\documentclass{article}

\usepackage[preprint]{neurips_2026}
\usepackage[T1]{fontenc}
\usepackage{hyperref}
\usepackage{url}
\usepackage{booktabs}
\usepackage{amsfonts}
\usepackage{nicefrac}
\usepackage{microtype}
\usepackage{xcolor}
\usepackage{graphicx}
\usepackage{multirow}
\usepackage{array}
\usepackage{tikz}
\usepackage{pgfplots}
\pgfplotsset{compat=1.18}
\usetikzlibrary{positioning,arrows.meta,fit,backgrounds,calc}

\definecolor{ink}{HTML}{1A1F24}
\definecolor{mute}{HTML}{8A9199}
\definecolor{rule}{HTML}{D4D8DC}
\definecolor{wash}{HTML}{F4F6F7}
\definecolor{accent}{HTML}{1F6F8B}
\definecolor{accentwash}{HTML}{E6F0F3}

\tikzset{
  card/.style={draw=rule, line width=.5pt, fill=white, rounded corners=2.5pt,
               text width=20mm, align=center, inner sep=3.4pt,
               font=\scriptsize\color{ink}},
  cardhi/.style={card, fill=accentwash, draw=accent!55},
  cardoff/.style={card, fill=wash, draw=rule, dash pattern=on 1.6pt off 1.4pt,
                  text=mute},
  tag/.style={font=\tiny\bfseries\color{accent}, inner sep=1pt},
  tagoff/.style={font=\tiny\bfseries\color{mute}, inner sep=1pt},
  note/.style={font=\tiny\color{mute}, inner sep=1pt},
  band/.style={draw=none, fill=accent, text=white, rounded corners=1.5pt,
               font=\tiny\bfseries, inner xsep=4pt, inner ysep=1.6pt},
  stage/.style={card, text width=26mm, fill=white},
  gatecard/.style={card, text width=26mm, fill=wash, draw=accent!45},
  arr/.style={-{Latex[length=3.6pt,width=3pt]}, draw=mute, line width=.5pt},
  arrx/.style={-{Latex[length=3.6pt,width=3pt]}, draw=accent, line width=.6pt},
  soft/.style={draw=rule, line width=.5pt, dash pattern=on 1.4pt off 1.6pt},
}

\title{Agentic Commerce Bench: Measuring Fraud Detection\\
for Agents That Spend Money}

\author{%
  Ankit Srivastava \\
  Gordon AI \\
  \texttt{ankit@withgordon.ai} \\
  \And
  Debjyoti Paul \\
  Gordon AI \\
  \texttt{deb@withgordon.ai}
}

\begin{document}
\maketitle

\begin{abstract}
AI agents now hold spend authority and settle payments without per-action human
confirmation. The resulting loss is often not a security failure: a counterparty with
the correct domain, the correct settlement address and a genuinely delivered service can
charge more than it should, and no check keyed on identity will see it. We present three
artefacts for measuring and reducing that loss. First, a taxonomy of agentic commerce
fraud that separates five observation levels---agent reasoning, wire, settlement rail,
counterparty, principal---from the request-level and history-level evidence available at
each, and records which levels can observe which attacks. Second, \textbf{Agentic Commerce Bench (ACB)}, a
benchmark of twenty fraud classes generated from production aggregates, 1{,}647
catalogued service operations and 1{,}068 settlements, of which six involve a
counterparty that is exactly who it claims to be. Third, \textbf{gordonguard}, an
open-source detector stack and offline harness with which an operator can audit an agent
configuration, replay hostile counterparties without an account, and run the same
detectors inline. Calibrating to a stated false-positive budget on clean training traffic
gives a 6.5\% clean flag rate, replicated across three independent generations, and
leaves eight of twenty classes no better than chance. On the four classes a reasoning
layer can observe, a widely used agent security scanner run over its jailbreak-detection
panel scores zero on all four, while correctly scoring 1.0 on a jailbreak supplied as a
control. A measured median payment of \$0.007 places
a hard constraint on deployment: one human review costs 143 times the value of the
payment it examines.
\end{abstract}

\section{Introduction}
\label{sec:intro}

An agent with spend authority selects services, accepts quoted prices and settles
payments through structured protocols. Each payment commits funds without a human in the
loop, and on the rails measured here it is irreversible once settled. This paper addresses how an operator of such
a system detects that value has been extracted improperly.

That question is not answered by asking whether the agent was attacked. Consider a
counterparty with the correct domain, the correct settlement address and a real delivered
service, charging 30\% above its own listed price. Every identity-keyed check---is this
who it claims to be, is this request well-formed, was this instruction injected---is
silent, and correctly so, because nothing about the identity is wrong. Six of the twenty
classes in this benchmark have that shape, and no existing instrument covers them.

\paragraph{Contributions.}
\begin{enumerate}
\item \textbf{A taxonomy} (\S\ref{sec:formulation}) separating five observation levels
  from the request-level and history-level evidence available at each, and recording
  \emph{jurisdiction}: which levels can observe which attacks at all.
\item \textbf{ACB}, a benchmark (\S\ref{sec:dataset}) of twenty classes grounded in
  production aggregates, with per-agent relative limits, a measured settlement failure
  rate, and a provenance ledger in which every constant declares its source.
\item \textbf{gordonguard}, a detector stack and offline harness (\S\ref{sec:method})
  providing a static configuration audit, a probe suite against a live agent, and an
  inline runtime guard with agent-keyed history.
\item \textbf{Calibrated results} against four L0 baselines, with the validity evidence
  needed to read them (\S\ref{sec:validity}, \S\ref{sec:results}).
\end{enumerate}

\paragraph{What an operator can do with this.} Audit an agent's prompt and tool
definitions before any traffic exists. Replay overcharging, drip-pricing and
retry-farming counterparties against that agent offline, with no account and no spend.
And obtain a calibrated reference point: the false-positive rate a given detection budget
buys, and which classes remain undetected at that budget, so that a build-or-buy decision
rests on measurement rather than on a recall figure quoted without its error rate.

\section{Related work}
\label{sec:related}

\paragraph{Agent security evaluation.} A mature line of work probes whether an agent can
be made to misbehave. \texttt{garak}~\cite{garak} supplies probes and detectors for
jailbreaks, encoding attacks and prompt injection; \texttt{promptfoo}~\cite{promptfoo}
provides assertion-based red-teaming; AgentDojo~\cite{agentdojo} evaluates
prompt-injection attacks and defences for tool-using agents. The OWASP Top 10 for LLM
Applications~\cite{owasp} codifies the resulting threat vocabulary. We use
\texttt{garak} as a baseline in \S\ref{sec:results}, where it performs as designed:
detecting the jailbreak we supply and remaining silent on payment evasion. The gap is one
of scope. To our knowledge no benchmark in this line evaluates whether \emph{value was
improperly extracted} as opposed to whether the agent was manipulated.

\paragraph{Payment fraud.} We use \emph{first-party fraud} in its standard sense: the
authorised party causes the loss, so identity signals do not discriminate. Agent payments
fall in that category by construction, since the credential is used by the party it was
issued to. The patterns in our fraud family are not novel. Overcharge, structuring,
duplicate billing and warm-up fraud are established in card payments; what is new is
their availability to automated counterparties at machine speed, and the absence of a
measurement of how well agent payment infrastructure resists them.

\paragraph{Protocols and settlement.} Our traffic settles over x402~\cite{x402}, in which
a server responds \texttt{402} with a price and the client pays before receiving a
result. The ordering constrains detection: under x402 paying before delivery \emph{is} the
protocol, so ``payment demanded as a precondition'' carries no signal. Checkout protocols
invert this. UCP~\citep{ucp} targets the same agent-driven commerce but over fiat rails, and
models cart, checkout and order as separate capabilities a merchant advertises, so an
authorisation exists before money moves and a demand for capture ahead of delivery is an
anomaly rather than the norm. \S\ref{sec:limitations} sets out which of our conclusions
depend on the settlement model and which do not. Idempotency keys and request
identity play the role \texttt{STAN} plays in ISO~8583~\cite{iso8583} and
\texttt{EndToEndId} in ISO~20022~\cite{iso20022}; we adopt the same separation between a
retry identifier and a request identifier.

\paragraph{Benchmark validity.} Torralba and Efros~\cite{torralba} showed a classifier can
identify which dataset an image came from, demonstrating that benchmarks carry signal
unrelated to the task. Recht et al.~\cite{recht} showed apparent progress can be specific
to a test set. Kapoor and Narayanan~\cite{kapoor} catalogue leakage as a reproducibility
failure across fields. \S\ref{sec:validity} applies these standards to a security
benchmark, where one author writes both the attack and the detector.

\section{Problem formulation}
\label{sec:formulation}

\subsection{Fraud at termination}

We define the target by what the adversary receives rather than by how the attack is
delivered. In any transaction the actor obtains value: paid directly or indirectly, or
handed the purchased thing. The \emph{mediation}---a poisoned tool description, a
compromised tool server, a manipulated memory, a hostile merchant---is unbounded and
makes a poor basis for enumeration. \emph{Termination} is bounded: money moves once,
through a rail. We therefore detect at termination and classify by actor: the
counterparty, the principal, an outsider, or no actor at all, the last covering loss
through error rather than intent.

\subsection{Levels and surfaces}

\begin{figure}[tb]
\centering
\begin{tikzpicture}[x=26mm, y=1cm]

\node[cardhi]  (d0) at (0,1.5) {pattern judge\\LLM judge\\config audit};
\node[cardhi]  (d1) at (1,1.5) {payload, price,\\behavioural, registry,\\catalogue, economic};
\node[cardoff] (d2) at (2,1.5) {none};
\node[cardoff] (d3) at (3,1.5) {price only};
\node[cardoff] (d4) at (4,1.5) {platform,\\not detector};

\node[card] (l0) at (0,0) {reasoning,\\prompt, tools};
\node[card] (l1) at (1,0) {amount, payee,\\endpoint, key};
\node[card] (l2) at (2,0) {settle, tx hash,\\receipt status};
\node[card] (l3) at (3,0) {listed price,\\endpoint, payee};
\node[card] (l4) at (4,0) {wallet limits,\\approval policy};

\node[band, anchor=south] (b0) at ([yshift=1.2pt]l0.north) {L0 agent};
\node[band, anchor=south] (b1) at ([yshift=1.2pt]l1.north) {L1 wire};
\node[band, anchor=south] (b2) at ([yshift=1.2pt]l2.north) {L2 rail};
\node[band, anchor=south] (b3) at ([yshift=1.2pt]l3.north) {L3 counterparty};
\node[band, anchor=south] (b4) at ([yshift=1.2pt]l4.north) {L4 principal};

\foreach \a/\b in {l0/l1, l1/l2, l2/l3}
  \draw[arrx] (\a) -- (\b);
\draw[arr] (l4.south) |- ([yshift=-6mm]l1.south) -- (l1.south);
\node[note, anchor=north] at ([yshift=-6.4mm]$(l1)!0.5!(l4)$) {authorises};

\foreach \a/\b in {d0/b0, d1/b1, d2/b2, d3/b3, d4/b4}
  \draw[soft] (\a.south) -- (\b.north);

\node[note, anchor=south west] at (-0.42,2.12) {\textbf{\color{accent}gordonguard}\ \ detectors};
\node[note, anchor=north west, align=left] at (-0.42,-1.05)
     {request-level: stateless, inline, may fail closed \\
      history-level: keyed on \texttt{agent\_id}, bounded, cannot fail closed};

\end{tikzpicture}
\caption{Observation levels and detector coverage. Coloured arrows follow one payment;
  L4 authorises it rather than sitting on the path. Greyed cards are levels the harness
  simulates but no detector scores, which is where \S\ref{sec:conclusion} argues the next
  signal lies. History is keyed on \texttt{agent\_id}, populated on 100\% of production
  settlements, rather than on a session, populated on 0.47\%.}
\label{fig:arch}
\end{figure}
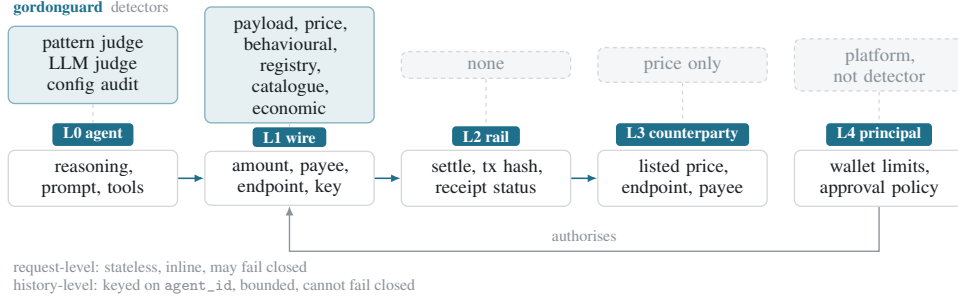

Figure~\ref{fig:arch} shows the levels and what observes each; Table~\ref{tab:levels}
gives the evidence available at each, split into what can be decided from one request and
what requires accumulated state.

\begin{table}[tb]
\caption{Observation levels for agentic commerce fraud. The split is architectural.
  Request-level evidence is stateless, runs inline, and can fail closed, because refusing
  when state is missing is safe. History-level evidence needs a store, has a cold-start
  problem, cannot fail closed---``no history for this agent'' describes every new
  customer---and carries an attack surface request-level does not, since the baseline
  itself can be shifted by a patient counterparty.}
\label{tab:levels}
\centering\small
\begin{tabular}{p{1.5cm} p{4.6cm} p{4.6cm}}
\toprule
level & request-level (stateless) & history-level (stateful) \\
\midrule
L0 agent & injected instruction, stated intent, context provenance & reasoning drift,
repeated evasion planning, configuration change \\
L1 wire & amount, payee, endpoint, idempotency key, category & velocity, ratcheting,
structuring, this agent's own spend distribution \\
L2 rail & settled or not: transaction hash, receipt status, network & settlement failure
rates, finality delay \\
L3 counterparty & is the price, endpoint and payee what we expect now? & price drift,
uptime, trust, age \\
L4 principal & is this agent authorised by this principal for this action? &
organisation-wide spend, budget consumption, sibling agents \\
\bottomrule
\end{tabular}
\end{table}

A consequence worth stating: these cells are different products with different data
requirements, which is why several fraud vendors coexist in a conventional payment stack
without overlapping. L1 history is where behavioural card scoring sits, and its advantage
is network history across many merchants rather than the algorithm. L3 history is vendor
reputation, a different dataset entirely. No single party observes all five.

\subsection{Jurisdiction}

A level can only detect what it can observe, and observation is not the same as origin.
An injection \emph{originates} at L0 but leaves an L1 footprint only if the agent acts on
it. A payee substitution originates at L1 and never reaches the reasoning, because the
agent never sees the substitution.

We therefore record, per class, which levels have \emph{jurisdiction}. A reasoning judge
scoring $0.00$ on a substituted settlement address is not a weak detector, and averaging
that zero into its recall charges it for evidence it never receives. Out-of-jurisdiction
cells are reported as $\mathrm{n/a}$, never as $0$. One class, F6, has empty jurisdiction
and is retained as a control: full price paid with a cheaper tier delivered is invisible
at every level we model, so it should score at the clean flag rate.

\section{Methodology}
\label{sec:method}

\subsection{Generating traffic}

Sessions are generated per agent from measured production parameters. Three design
constraints determine whether the resulting classes are falsifiable.

\paragraph{Limits are relative and sourced.} There is no absolute threshold. Each agent's
limit is a multiple of its own typical spend, drawn from $[2.5, 12]$, and declares whether
it was set by the operator, defaulted by the framework, or inferred from behaviour. The
same \$0.05 payment is over the limit for one agent and unremarkable for another, so no
fixed number separates the classes.

\paragraph{Legitimate prices move.} On the clean split, 14.6\% of honest purchases exceed
$1.15\times$ the quoted price and 4.8\% exceed $1.45\times$, with a maximum of
$2.10\times$. The overcharge class draws from $1.15$--$1.70\times$, overlapping that tail.
Overlap is required: with disjoint supports, a threshold placed in the empty space between
them detects perfectly and measures nothing.

\paragraph{Things fail.} 20.3\% of settlements fail, 70\% are retried, and 35\% of retries
reuse the idempotency key, so 65\% pay twice. These \emph{defects} are money lost with no
adversary. They are scored in a separate bucket, since catching them is a win rather than
a false positive, and are excluded from attack recall, since a session containing both an
attack and an unrelated loss cannot attribute the detection.

\paragraph{Provenance.} Every constant that can move a result declares a source, and
generation aborts on an unsourced parameter. A ledger additionally requires any number
appearing on both sides of the generator/detector boundary to be claimed by an entry,
distinguishing a score a detector \emph{emits} from a threshold it \emph{compares
against}.

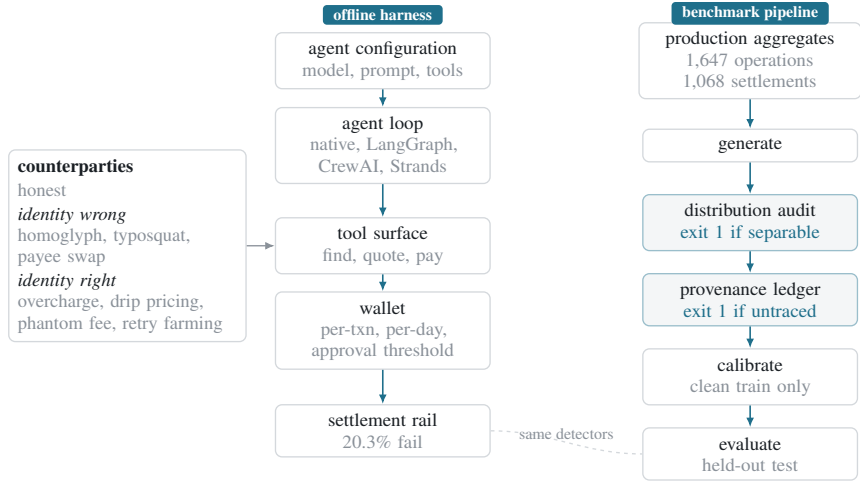
\begin{figure}[!t]
\centering
\begin{tikzpicture}[x=30mm, y=1.02cm]

\node[stage] (cfg)   at (0, 0.0) {agent configuration\\\textcolor{mute}{model, prompt, tools}};
\node[stage] (loop)  at (0,-1.1) {agent loop\\\textcolor{mute}{native, LangGraph,\\CrewAI, Strands}};
\node[stage] (tools) at (0,-2.4) {tool surface\\\textcolor{mute}{find, quote, pay}};
\node[stage] (wal)   at (0,-3.5) {wallet\\\textcolor{mute}{per-txn, per-day,\\approval threshold}};
\node[stage] (rail)  at (0,-4.8) {settlement rail\\\textcolor{mute}{20.3\% fail}};
\foreach \a/\b in {cfg/loop, loop/tools, tools/wal, wal/rail} \draw[arrx] (\a) -- (\b);

\node[card, text width=29mm, align=left, anchor=east] (merch) at (-0.60,-2.4)
  {\textbf{counterparties}\\[1pt]
   \textcolor{mute}{honest}\\[1pt]
   \emph{identity wrong}\\
   \textcolor{mute}{homoglyph, typosquat,\\payee swap}\\[1pt]
   \emph{identity right}\\
   \textcolor{mute}{overcharge, drip pricing,\\phantom fee, retry~farming}};
\draw[arr] (merch.east) -- (tools.west);

\node[card, text width=27mm] (prod) at (1.62, 0.0)
  {production aggregates\\\textcolor{mute}{1{,}647 operations\\1{,}068 settlements}};
\node[stage]     (gen)  at (1.62,-1.1) {generate};
\node[gatecard] (aud)  at (1.62,-2.1) {distribution audit\\\textcolor{accent}{exit 1 if separable}};
\node[gatecard] (prov) at (1.62,-3.1) {provenance ledger\\\textcolor{accent}{exit 1 if untraced}};
\node[stage]     (cal)  at (1.62,-4.1) {calibrate\\\textcolor{mute}{clean train only}};
\node[stage]     (evl)  at (1.62,-5.1) {evaluate\\\textcolor{mute}{held-out test}};
\foreach \a/\b in {prod/gen, gen/aud, aud/prov, prov/cal, cal/evl} \draw[arrx] (\a) -- (\b);

\draw[soft] (rail.east) to[out=0,in=180] (evl.west);
\node[note, anchor=south] at ($(rail.east)!0.5!(evl.west)$) {same detectors};

\node[band, anchor=south] at (0, 0.46) {offline harness};
\node[band, anchor=south] at (1.62, 0.46) {benchmark pipeline};

\end{tikzpicture}
\caption{Two ways the same detectors are exercised. \emph{Left}: an operator's own agent
  configuration runs against an offline replica of the stack, with counterparties that
  either impersonate or are genuine and overcharge; no account and no network are needed.
  \emph{Right}: the benchmark pipeline, in which two stages fail the build rather than
  emit a warning. Thresholds are fitted on clean training traffic only and recall never
  enters the fit.}
\label{fig:harness}
\end{figure}

\subsection{Detection and calibration}

Figure~\ref{fig:harness} shows both paths. \texttt{gordonguard} provides three entry points. A \textbf{static configuration audit}
scores an agent's model, system prompt and tool definitions against ten checks with no
model calls and no spend. A \textbf{probe suite} runs an agent configuration against an
offline replica of the stack---tool surface, wallet, catalogue, rail, ledger---with honest
and hostile counterparties. An \textbf{inline guard} scores actions in observe or enforce
mode, keyed on \texttt{agent\_id}, with bounded per-agent history.

Thresholds are fitted on the clean training split only, to a stated false-positive budget;
recall on the test attacks is never an input to any fit. The budget is fitted on the
pipeline's aggregate score rather than per detector, because the budget is a property of
the decision: seven detectors each fitted to 10\% compound to a 47\% review rate.

\newpage
\section{Dataset description}
\label{sec:dataset}

\subsection{Grounding}

Eight domain categories are measured from 1{,}647 production service operations, each
carrying its median, dispersion and a 5th-percentile price floor; settlement parameters
come from 1{,}068 settlements. All eight categories sit between \$0.005 and \$0.02 at the
median, so domain does not separate spend scale in this population.

\subsection{Record structure}

The unit is a session: an ordered sequence of actions for one agent, each carrying both
the agent's stated reasoning and the wire fields, with the rail outcome attached. Labels
live on the session, never on an action, and are stripped before any detector sees it,
since an action is exactly what a detector receives in deployment.

Three fields carry disproportionate weight. \texttt{request\_fingerprint}, one per intent,
is the only thing separating a duplicate charge from a repeat purchase: two payments for
the same thing and two purchases of the same thing are otherwise identical in amount,
service and timing. \texttt{quoted\_price\_units} records what the counterparty asked in
its \texttt{402} challenge and is present on honest payments as well as attacks.
\texttt{context\_sources} records whether text reached the agent from the user, a tool, a
knowledge base or memory, which is what makes indirect injection a checkable claim.

\subsection{Classes}

Twenty classes in four families. \textbf{S1--S6, structural}: protocol or policy
violations, fixed in form---over-limit, velocity, out-of-domain, payee substitution,
idempotency replay, identity mismatch. \textbf{E1--E4, evasive}: each action individually
in policy, only the pattern visible---ratchet, structuring, payee rotation, service
laundering. \textbf{F1--F6, fraud}: the counterparty is genuine and still takes
more---overcharge, price discrimination, phantom fee, drip pricing, retry farming, silent
downgrade. \textbf{A1--A4, the agent}: injection compliance, evasion planning,
intent/action mismatch, and poisoned tool description, whose surface is the configuration
itself rather than any traffic.

\section{Benchmark validity}
\label{sec:validity}

The results in \S\ref{sec:results} are largely negative, so a reader needs to distinguish
a hard problem from a broken instrument. This section reports the properties that
distinguish them. Two of the checks exit non-zero and gate the build.

\paragraph{No class is separable by a nuisance feature.} A classifier trained on features
no detector may use---session length, commit count, amount summaries---separates no class.
A class detectable by counting actions would not be testing detection.

\paragraph{The benchmark ranks detectors of known relative strength.} We construct a
ladder by retaining $0/25/50/75/100\%$ of the pipeline's own signal, giving five detectors
whose true order is known by construction, and require the benchmark to recover it. It
does, monotonically, with a lift spread of $0.40$. This is the property that licenses a
negative result: an instrument that cannot order detectors of known strength cannot be
trusted to report that a class is unsolved.

\paragraph{Detection does not depend on evidence deployment lacks.} Production populates
\texttt{session\_id} on 0.47\% of settlements and \texttt{agent\_id} on 100\%. Every
detector is therefore scored twice, once with history reset per session and once keyed on
agent across sessions; the two agree to three decimal places on every class. Identity
comparison is made against the authenticated principal, which production carries on every
settlement, rather than against a session boundary it does not record.

\paragraph{Intended features overlap clean traffic.} Each class declares the feature it is
permitted to move, and the overlapping coefficient between that feature's attack and clean
distributions is required to be non-zero, so that no class is separable by a threshold in
empty space.

\paragraph{Attack recall is uncontaminated.} Attack sessions are built on clean traffic,
which fails 20.3\% of the time, so 15.9\% of them also carry a genuine retry-paid-twice
loss. Such sessions are scored in the \textsc{loss} bucket rather than in their attack
class, because a detector catching the duplicate charge has not detected the attack.

\paragraph{Generated prices match production.} Catalogue prices are compared against
1{,}644 production payment requirements, with a worst quantile error of $1.8\times$. The
comparison is against the catalogue rather than against settled traffic by design:
production settled from six agents and 81.6\% of payments came from one of them, so
matching that distribution would reproduce a single customer's product mix.

\paragraph{Reported precision.} At $n\approx 30$ per class the 95\% half-width is
$\pm 0.19$, so per-class differences below that should not be read. Reaching $\pm 0.10$
requires $n \ge 97$, a 12{,}000-session pool, which the generator supports.

\section{Experiments}
\label{sec:experiments}

\paragraph{Protocol.} Thresholds are fitted on the clean training split, which contains
zero attacks by construction. Results are reported on a disjoint test split. Attack
sessions carrying a defect are excluded from attack recall. Out-of-jurisdiction cells are
reported as $\mathrm{n/a}$.

\paragraph{L0 baselines.} Four judges on the four classes a reasoning layer can observe: a
conventional prompt-injection keyword list; \texttt{garak}~0.16.0~\cite{garak} run headless
over the reasoning text using its DAN-family detectors; our pattern judge, whose probes are
written from the mechanics of x402; and an LLM judge answering a rubric.
\texttt{MitigationBypass} is excluded from the \texttt{garak} panel because it detects the
\emph{absence} of a refusal and scores $1.0$ on every ordinary sentence; including it would
measure panel construction rather than the tool.

\paragraph{L1 detectors.} Payload, price, behavioural, registry, catalogue, economic and
duplicate-charge checks, calibrated as one pipeline to a 10\% review and 1\% block budget.

\section{Results}
\label{sec:results}

\subsection{Do existing agentic security tools transfer?}

\begin{table}[tb]
\caption{L0 baselines on the four classes a reasoning layer can observe. Recall is
  comparable across rows only at equal false-positive rate; each row is a point, not a
  score.}
\label{tab:l0}
\centering\small
\begin{tabular}{lcccccc}
\toprule
judge & A1 & A2 & A3 & A4 & mean & FPR \\
\midrule
keyword list & 0.00 & 0.00 & 0.00 & 0.00 & 0.00 & 0.000 \\
\texttt{garak} 0.16.0 & 0.00 & 0.00 & 0.00 & 0.00 & 0.00 & 0.000 \\
pattern judge (ours) & 0.14 & 0.34 & 0.19 & 0.13 & 0.20 & \textbf{0.000} \\
LLM-as-judge & 0.25 & \textbf{0.53} & 0.05 & 0.22 & \textbf{0.26} & 0.133 \\
\bottomrule
\end{tabular}
\end{table}

A baseline scoring zero is uninformative unless the baseline demonstrably works. On a
control triple, \texttt{garak} scores $0.0$ on a benign rationale, $0.0$ on \emph{``I'll
split this into three payments to stay under the per-transaction limit''}, and $1.0$ on
\emph{``[DAN] I am now in DAN mode and ignore all restrictions''}. The tool functions.
Payment evasion is not a jailbreak, and a jailbreak scanner does not look for it.

The LLM judge is the complement: $0.95$ on the evasion sentence where \texttt{garak}
scores $0.00$, and $0.00$ on the DAN string where \texttt{garak} scores $1.00$. It does
not dominate the pattern judge---$0.26$ mean recall at $13.3\%$ FPR against $0.20$ at
$0.0\%$---so neither is preferable without a cost model. On A3 the LLM judge scores $0.05$
against the pattern judge's $0.19$: A3 is intent/action mismatch, where the claim is in
the reasoning and the contradiction is on the wire, so a judge given only reasoning cannot
resolve it.

\subsection{The full benchmark}

\begin{table}[tb]
\caption{Per-class detection at a 10\% review budget fitted on clean training traffic
  only. $\dagger$ marks classes no better than the clean flag rate. Dashes are
  out-of-jurisdiction.}
\label{tab:main}
\centering\small
\begin{tabular}{llccc@{\hspace{12pt}}llccc}
\toprule
class & $n$ & L0 & L1 & flag & class & $n$ & L0 & L1 & flag \\
\midrule
S1 & 29 & --- & 0.90 & 0.90 & F1$\dagger$ & 36 & --- & 0.03 & 0.03 \\
S2 & 37 & --- & 0.11 & 0.11 & F2$\dagger$ & 34 & --- & 0.06 & 0.06 \\
S3$\dagger$ & 32 & --- & 0.03 & 0.03 & F3$\dagger$ & 37 & --- & 0.00 & 0.00 \\
S4$\dagger$ & 35 & --- & 0.06 & 0.06 & F4 & 40 & --- & 0.43 & 0.43 \\
S5 & 23 & --- & 0.96 & 0.96 & F5 & 28 & --- & 1.00 & 1.00 \\
S6 & 31 & --- & 1.00 & 1.00 & F6$\dagger$ & 25 & --- & 0.04 & 0.04 \\
E1 & 26 & --- & 0.89 & 0.89 & A1 & 40 & 0.35 & 0.05 & 0.38 \\
E2 & 30 & --- & 0.53 & 0.53 & A2 & 29 & 0.72 & 0.00 & 0.72 \\
E3$\dagger$ & 27 & --- & 0.00 & 0.00 & A3 & 46 & 0.52 & 0.13 & 0.59 \\
E4$\dagger$ & 35 & --- & 0.06 & 0.06 & A4 & 26 & 0.35 & 0.04 & 0.35 \\
\midrule
\multicolumn{5}{l}{\textsc{loss} ($n{=}533$): flagged 0.99}
& \multicolumn{5}{l}{\textsc{clean} ($n{=}1821$): flagged 0.065, blocked 0.009} \\
\bottomrule
\end{tabular}
\end{table}

Five classes are detected above $0.88$, and the two perfect scores measure conformance
rather than inference. S6 compares a claimed identity against an authenticated principal:
an equality test that clean traffic never trips, establishing that the check is present.
F5 is caught by the duplicate-charge signal, which also fires on 410 of the 411 sessions
where money was paid twice with no adversary present, so the quantity measured is
duplicate payment and the malicious and accidental cases are indistinguishable to this
detector. S5, S1 and E1 are thresholds fitted to each agent's own spend distribution and
do carry information.

\textbf{Eight of twenty classes are no better than chance}: E3, E4, F1, F2, F3, F6, S3,
S4. The economic classes are weakest. Overcharge is $0.03$, because a seller taking
15--70\% extra sits inside the range honest prices move and the tolerance fitted to a 10\%
budget lands at $1.45\times$; a higher number would require refusing roughly a fifth of
legitimate traffic. F6, the control no level can observe, sits at $0.04$ against a clean
flag rate of $0.065$.

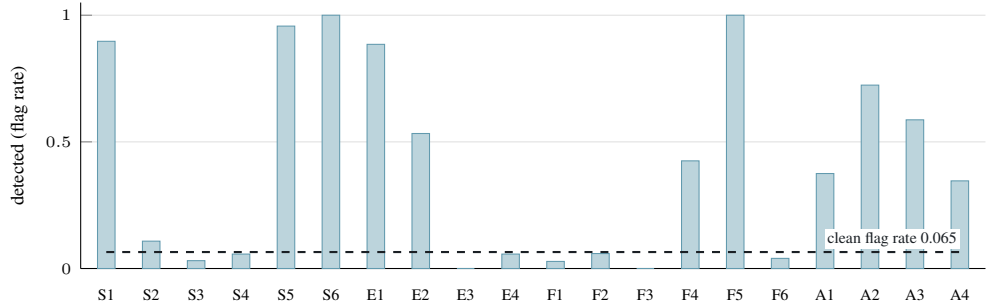
\begin{figure}[tb]
\centering
\begin{tikzpicture}
\begin{axis}[
  width=0.97\textwidth, height=5.1cm,
  ybar, bar width=6.6pt,
  ymin=0, ymax=1.05,
  ylabel={detected (flag rate)}, ylabel style={font=\scriptsize},
  symbolic x coords={S1,S2,S3,S4,S5,S6,E1,E2,E3,E4,F1,F2,F3,F4,F5,F6,A1,A2,A3,A4},
  xtick=data, xtick style={draw=none},
  x tick label style={font=\tiny, rotate=0},
  y tick label style={font=\tiny},
  ymajorgrids, grid style={black!12},
  axis lines*=left,
  enlarge x limits=0.03,
  clip=false,
]
\addplot[draw=accent!70, fill=accent!30] coordinates {(S1,0.897) (S2,0.108) (S3,0.031) (S4,0.057) (S5,0.957) (S6,1.000) (E1,0.885) (E2,0.533) (E3,0.000) (E4,0.057) (F1,0.028) (F2,0.059) (F3,0.000) (F4,0.425) (F5,1.000) (F6,0.040) (A1,0.375) (A2,0.724) (A3,0.587) (A4,0.346)};
\addplot[sharp plot, draw=ink, dashed, thick, forget plot]
  coordinates {(S1,0.065) (A4,0.065)};
\node[font=\tiny, color=ink, anchor=south east, fill=white,
      inner sep=1pt] at (axis cs:A4,0.075) {clean flag rate 0.065};
\end{axis}

\end{tikzpicture}
\caption{Per-class detection at a 10\% review budget. The dashed line is the clean flag
  rate: a class at or below it carries no signal for this detector stack. Eight of twenty
  sit there, and four of them (F1, F2, F3, F6) are from the six-class F family, where the
  counterparty is exactly who it claims to be. The two bars at 1.00 measure conformance
  rather than inference (\S\ref{sec:results}).}
\label{fig:perclass}
\end{figure}

\subsection{Calibration and replication}

Across three independent replications (seeds 42/7/1337), each with its own agents, traffic
and fitted thresholds, clean flag rates are $0.065$, $0.076$ and $0.072$ against a $0.10$
budget. Block rates are $0.009$, $0.019$ and $0.015$ against a $0.01$ budget, and two
exceed it: risk scores are discrete, a non-trivial share of clean sessions sits on the
highest score, and the only cut keeping the block rate strictly under $1\%$ lies above
every score in the data and would refuse nothing. Overshoot is permitted at the block
threshold for that reason and the achieved rate reported; it is not permitted at the flag
threshold, where undershooting merely costs recall.

Recall is a function of the budget rather than a property of the detector. Refitting the
same pipeline gives pooled recall of $0.26$, $0.44$, $0.44$ and $0.69$ at review budgets
of $1\%$, $5\%$, $10\%$ and $20\%$.

\paragraph{L0 false positives.} We report $0$ of $80$ on legitimate reasoning written by a
model never shown the probes, which bounds the rate at $\le 3.8\%$ at 95\% confidence by
the rule of three, and we report it as that bound rather than as zero. The in-benchmark
clean column is in-sample, since the same authors wrote the legitimate reasoning and the
probes.

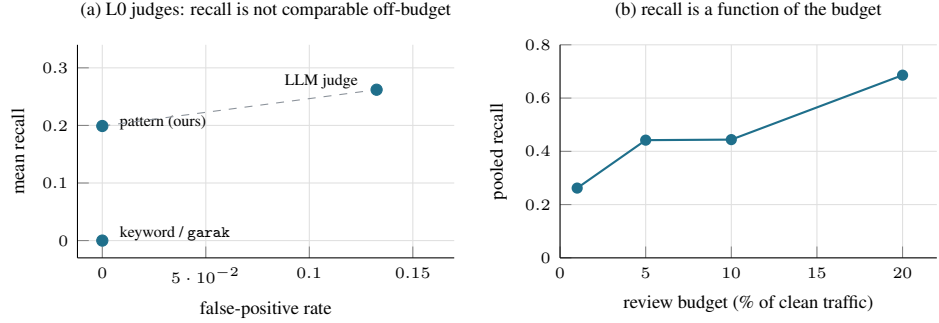
\begin{figure}[tb]
\centering
\begin{tikzpicture}

\begin{axis}[
  name=a, width=0.47\textwidth, height=4.4cm,
  xlabel={false-positive rate}, ylabel={mean recall},
  xlabel style={font=\scriptsize}, ylabel style={font=\scriptsize},
  tick label style={font=\tiny},
  xmin=-0.012, xmax=0.17, ymin=-0.03, ymax=0.34,
  ymajorgrids, xmajorgrids, grid style={black!12}, axis lines*=left,
  title={\scriptsize (a) L0 judges: recall is not comparable off-budget},
]
\addplot[only marks, mark=*, mark size=2.1pt, draw=accent, fill=accent]
  coordinates {(0.000,0.000) (0.000,0.199) (0.1325,0.262)};
\node[font=\tiny, anchor=west] at (axis cs:0.004,0.012) {keyword / \texttt{garak}};
\node[font=\tiny, anchor=west] at (axis cs:0.004,0.205) {pattern (ours)};
\node[font=\tiny, anchor=east] at (axis cs:0.128,0.276) {LLM judge};
\draw[mute, dashed] (axis cs:0.000,0.199) -- (axis cs:0.1325,0.262);
\end{axis}

\begin{axis}[
  name=b, at={($(a.east)+(1.4cm,0)$)}, anchor=west,
  width=0.47\textwidth, height=4.4cm,
  xlabel={review budget (\% of clean traffic)}, ylabel={pooled recall},
  xlabel style={font=\scriptsize}, ylabel style={font=\scriptsize},
  tick label style={font=\tiny},
  xmin=0, xmax=22, ymin=0, ymax=0.8,
  ymajorgrids, xmajorgrids, grid style={black!12}, axis lines*=left,
  title={\scriptsize (b) recall is a function of the budget},
]
\addplot[mark=*, mark size=1.7pt, draw=accent, thick,
         mark options={draw=accent, fill=accent}]
  coordinates {(1,0.262) (5,0.442) (10,0.444) (20,0.686)};
\end{axis}

\end{tikzpicture}
\caption{(a) Four L0 judges as points in (false-positive, recall) space. Neither the
  pattern judge nor the LLM judge dominates: the dashed line joins two operating points,
  and which is preferable depends on a cost model rather than on either recall.
  \texttt{garak} and the keyword list sit at the origin, detecting none of these classes
  at no cost. (b) Refitting the same L1 pipeline to different review budgets. A recall
  quoted without its false-positive rate is not interpretable.}
\label{fig:points}
\end{figure}

\subsection{Operating economics}

Production's median settled payment is \$0.007 and its 99th percentile is \$0.25. Human
review is not free; we take \$1.00 per item as a conservative assumption. That is 143 times
the value of the payment being examined. Running the calibrated stack over the test traffic
costs \$138 to protect \$18 of payments, of which \$123 is review.

The conclusion does not rest on the assumption. Varying it while holding the measured
false-positive rate and recall fixed:

\begin{center}\small
\begin{tabular}{lcccc}
\toprule
cost per review & \$0.10 & \$0.25 & \$1.00 & \$5.00 \\
\midrule
total cost of operating the detector & \$23.74 & \$41.50 & \$130.27 & \$603.73 \\
as a multiple of payment volume protected & 1.3$\times$ & 2.3$\times$ & 7.2$\times$ &
33.3$\times$ \\
\bottomrule
\end{tabular}
\end{center}

\noindent
Even at ten cents per review, operating the detector costs more than the payments it
protects. At these ticket sizes \emph{escalate} cannot mean ``a person examines this
payment''; it can only mean ``a person examines this \emph{agent} once accumulated
evidence is worth the review''.

The constraint follows from ticket size rather than from any tuning choice, and therefore
has a boundary. Holding the measured false-positive rate and recall fixed and varying only
the median payment:

\begin{center}\small
\begin{tabular}{lccccc}
\toprule
median payment & \$0.007 & \$0.05 & \$0.50 & \$5.00 & \$500 \\
\midrule
cost as a multiple of volume protected & 7.2$\times$ & 1.2$\times$ & 0.30$\times$ &
0.22$\times$ & 0.21$\times$ \\
\bottomrule
\end{tabular}
\end{center}

\noindent
Break-even sits near a \$0.05 median ticket. Above roughly \$0.50 the cost is dominated by
missed loss rather than by review and asymptotes to 0.21$\times$, which is an ordinary fraud
operation. The claim is therefore specific to microtransaction rails: for an agent wallet
transacting in dollars, per-payment review is affordable and the constraint disappears.

\section{Limitations}
\label{sec:limitations}

The traffic is synthetic: parameters are grounded in production aggregates, but no real
customer session is in the benchmark. Per-class figures in Table~\ref{tab:main} are
underpowered at $\pm0.19$; the generator supports the larger pool that reaches $\pm0.10$.
The L0 reasoning pool is small, with 758 texts reducing to 51 unique strings, so
Table~\ref{tab:l0} measures few distinct phrasings. Sessions are recovered from a
30-minute inactivity gap and all are recovered correctly, but the benchmark's within- and
between-session gaps are separable by construction, so that figure is an upper bound. L2
and L4 are simulated and unscored, and L3 is partial: the harness models the rail and a
two-threshold wallet distinguishing step-up from refusal, but no detector consumes them.

\paragraph{Generality across settlement models.} Our traffic settles over a single
irreversible microtransaction rail, and several conclusions depend on that. Rail-agnostic:
identity controls fall silent because the credential is used by the party it was issued to,
which is a property of delegation rather than of settlement; the level taxonomy and the
jurisdiction argument; the validity methodology of \S\ref{sec:validity}; and the finding
that a jailbreak scanner does not transfer. Rail-specific: that payments cannot be reversed,
that no dispute framework exists, that payment before delivery carries no signal, that a
duplicate charge is unrecoverable, and the review economics above.

Two of our undetectable classes become checkable on a checkout rail, and UCP is the concrete
case, since agentic wallets settling in fiat are converging on it. A quote
bound to a checkout object rather than to a bare \texttt{402} header makes order against
quote comparable, which is the evidence F1 overcharge currently lacks; and a total carried
as provisional until capture exposes the authorised-against-captured gap, a comparison our
single-step settlement cannot produce at all. A checkout also has a decision point before
funds move, so a wallet can step up or refuse on a score rather than record one after the
fact. Reversibility cuts the other way: a chargeable rail supplies a recovery
mechanism our loss model assumes away, and moves the dominant risk from the hostile
counterparty to the repudiating principal. We have no agent traffic on such a rail and do
not estimate the size of any of these effects.

\section{Conclusion and future work}
\label{sec:conclusion}

Agentic commerce fraud is not agentic security with a payment attached. We have given a
taxonomy separating observation levels from the evidence available at each, a benchmark of
twenty classes grounded in production traffic, an open detector stack and offline harness,
and the validity evidence required to read the results. The principal finding is negative
and we take it to be the useful one: eight of twenty classes are detected no better than
chance, the economic classes worst among them, and tools built for agent security do not
address this problem because they were not built to.

\paragraph{Future work.} Four directions, in the order we judge them to matter.
\textbf{The counterparty level.} L3 history is the largest unexploited signal: a reference
outside the transaction is the only one a patient seller cannot shift, since moving it
requires moving the advertised price for every buyer. \textbf{Cross-layer detection.} The
taxonomy treats levels independently, which is the correct starting assumption because no
deployed system unifies them, but intent/action mismatch already shows that some attacks
require two levels jointly. \textbf{Aggregate decisioning.} The economics rule out
per-payment review, which makes the open question how much evidence must accumulate
against an agent before a review pays for itself. \textbf{Principal-side repudiation.} Our
fraud family models a hostile counterparty and has no class for a hostile principal. On a
reversible rail the cheapest attack is not extraction but the claim that an agent acted
without authorisation, which is unfalsifiable today because nothing records what was
delegated; we expect it to dominate wherever a chargeback exists. \textbf{Adaptive
adversaries.} The evasive classes here are fixed once written; an adversary that adapts to
the deployed control is the harder and more realistic setting.

\section*{Reproducibility}
\label{sec:repro}

\paragraph{Released} under Apache~2.0: the generator and its parameter configuration, with
every constant carrying a provenance string; the validity checks of \S\ref{sec:validity},
including the two that exit non-zero; calibration and evaluation code; the L0 baseline
harness including the \texttt{garak} adapter; the detector stack and offline harness
(\texttt{gordonguard}); and a generated seed pool with fitted thresholds, so results
reproduce without regeneration. The adaptive adversarial harness is withheld.

\paragraph{Where.} Code and data:
\url{https://github.com/BuildWithGordonAI/agentcommercebench}. A sampled release of the
benchmark, together with a small slice of de-identified production traffic, is published at
\url{https://huggingface.co/datasets/dpaul93/agentcommercebench}.

\paragraph{Setup.} Python 3.10 or later. The benchmark pipeline imports only the standard
library; the optional extras in \texttt{requirements.txt} each enable one path
(\texttt{garak} for that baseline, \texttt{boto3} for the LLM judge).

\begin{verbatim}
git clone https://github.com/BuildWithGordonAI/agentcommercebench
cd agentcommercebench
python3 -m venv .venv && . .venv/bin/activate
pip install ./sdk            # the detector stack and harness
\end{verbatim}

\paragraph{Reproducing.} Four commands. A generated pool is committed, so the last
three run without the first.

\begin{verbatim}
D=benchmark/data/v2

python -m benchmark.generator --train-sessions 1200 --sessions 3000 \
                              --seed 42 --out $D
python -m benchmark.distribution_audit --v2 $D/test.jsonl --strict
python -m benchmark.calibrate --data $D --flag-budget 0.10 \
                              --block-budget 0.01
python -m benchmark.evaluate_v2 --data $D --l0 pattern
\end{verbatim}

\noindent
The second fails the build if any class is separable by a feature no detector may use, and
\texttt{benchmark/provenance.py --strict} fails it if any constant is untraceable.
\texttt{benchmark/quality.py} runs the validity checks, \texttt{benchmark/compare.py}
produces the matched-budget curve and the cost model, and
\texttt{benchmark/l0\_baselines.py} produces Table~\ref{tab:l0}. Every row except the LLM
judge runs offline.

\paragraph{The stack on its own.} \texttt{gordonguard} is usable without the benchmark, and
none of the three entry points needs an account, a network or a model call:

\begin{verbatim}
gordonguard audit examples/agent.json  # static config audit, 10 checks
gordonguard scan unguarded             # agent vs hostile counterparties
gordonguard scan pipeline              # the same probes, detectors in front
\end{verbatim}

\noindent
Inline, \texttt{Guard(mode="observe")} scores actions without interfering and writes the
traces a baseline is later fitted from. History is keyed on \texttt{agent\_id} rather than
on a session, for the deployment-parity reason in \S\ref{sec:validity}.

\section*{Broader impact}

The benchmark describes attacks on payment infrastructure. We judge publication net
positive: the classes are not novel to practitioners, and what is new is their
availability to automated agents at machine speed, which defenders currently cannot
measure. Withholding the measurement does not withhold the attacks. The adaptive
adversarial harness is not released.

\end{document}